\documentclass{iaus}
\usepackage{graphicx}
\usepackage{sidecap}

\newcommand{\EQ}{\begin{equation}}
\newcommand{\EN}{\end{equation}}
\newcommand{\EQA}{\begin{eqnarray}}
\newcommand{\ENA}{\end{eqnarray}}

\newcommand{\EEqs}[2]{Equations~(\ref{#1}) and~(\ref{#2})}

\newcommand{\Fig}[1]{Fig.~\ref{#1}}

\newcommand{\bra}[1]{\langle #1\rangle}

{}
{}
{}

{}
{}
{}
{}
{}
{}
{}
{}
{}
{}
{}
{}
{}
{}
{}
{}
{}
{}

{}
{}
{}

{}

{}
{}

\newcommand{\rrr}{\hat{\mbox{\boldmath $r$}} {}}

\newcommand{\gggg}{\mbox{\boldmath $g$} {}}

\newcommand{\UU}{\mbox{\boldmath $U$} {}}

\newcommand{\BB}{\mbox{\boldmath $B$} {}}

\newcommand{\JJ}{\mbox{\boldmath $J$} {}}

\newcommand{\AAA}{\mbox{\boldmath $A$} {}}

\newcommand{\ff}{\mbox{\boldmath $f$} {}}

\newcommand{\FF}{\mbox{\boldmath $F$} {}}

\newcommand{\nab}{\mbox{\boldmath $\nabla$} {}}

\newcommand{\SSSS}{\mbox{\boldmath ${\sf S}$} {}}

\newcommand{\erf}{{\rm erf}}

\newcommand{\DD}{{\rm D} {}}

\newcommand{\const}{{\rm const}  {}}

\def\cs{c_{\rm s}}

\def\kf{k_{\rm f}}

\def\urms{u_{\rm rms}}

\def\half{{\textstyle{1\over2}}}

\def\onethird{{\textstyle{1\over3}}}

\newcommand{\yapj}[3]{ #1, {ApJ,} {#2}, #3}

\newcommand{\yapjl}[3]{ #1, {ApJ,} {#2}, #3}

\newcommand{\yana}[3]{ #1, {A\&A,} {#2}, #3}

\newcommand{\ygrl}[3]{ #1, {Geophys.\ Res.\ Lett.,} {#2}, #3}

\newcommand{\ymn}[3]{ #1, {MNRAS,} {#2}, #3}

\title[Plasmoid ejections driven by dynamo action] 
{Plasmoid ejections driven by dynamo action underneath a spherical surface}

\author[J{\"o}rn Warnecke, Axel Brandenburg, Dhrubaditya Mitra]   
{J{\"o}rn Warnecke$^{1,2}$
 \and Axel Brandenburg$^{1,2}$
\and Dhrubaditya Mitra$^{1}$}
\affiliation{$^1$Nordita, AlbaNova University Center, \\Roslagstullsbacken 23,
SE-10691 Stockholm, Sweden \\email: {\tt joern@nordita.org} \\[\affilskip]
$^2$Department of Astronomy, AlbaNova University Center, \\Stockholm University, 
SE 10691 Stockholm, Sweden}

\pubyear{2011}
\volume{274}  
\pagerange{119--126}
\jname{Advances in Plasma Astrophysics}
\editors{A.Bonanno,  E. de Gouveia dal Pino and A. Kosovichev.}
\begin{document}

\maketitle

\begin{abstract}
We present a unified three-dimensional model of the convection zone
and upper atmosphere of the Sun in spherical geometry.
In this model, magnetic fields, generated by a 
helically forced dynamo in the convection zone,
emerge without the assistance of magnetic buoyancy.
We use an isothermal equation of state with gravity and density 
stratification.
Recurrent plasmoid ejections, which rise through the outer atmosphere,
is observed. 
In addition, the current helicity of the small--scale field is
transported outwards and form large structures like magnetic clouds.

\end{abstract}
\firstsection 
\section{Introduction}
Usually, magnetic phenomena in the atmosphere of the Sun,
e.g., the formation of active regions and sunspots, emergence of magnetic fields, and
coronal mass ejections, are described in terms of magnetic flux tubes.
These flux tubes are supposed to form at the tachocline, 
rise through the convection zone almost undeformed by the effects
of magnetic buoyancy and reach into the photosphere by creating
bipolar regions such as sunspots. Above the photosphere twisted
magnetic fields are observed to form arch-like structure. 
However, there is no clear evidence for the existence of magnetic
flux tubes inside the deep convection zone.
Direct numerical simulations of large-scale dynamos suggest that
flux tubes are primarily a
feature of the kinematic regime, but tend to be less pronounced in
the nonlinear stage (K\"apyl\"a et al.\ 2008).
In addition, magnetic buoyancy, which is usually believed to be the driver of
flux tube emergence, can be compensated sufficiently by downward pumping
resulting from the stratification of turbulence intensity
in the solar convection zone (Nordlund et al.\ 1992; Tobias et al.\ 1998).

In an earlier work (Warnecke \& Brandenburg 2010) we have suggested
an alternative approach. 
This is a two-layered approach, where the simulation of a turbulent
large-scale dynamo in the convection zone is coupled to a
simplified solar atmosphere model. Magnetic buoyancy does not play
an important role in this model. Such a simplified model,
in Cartesian coordinates, was able to capture some of the
observed qualitative features. In this work we extend this model
further, by going to a spherical coordinate system,
in the presence of density stratification and gravity. 
Here we present preliminary results from such a study. 
We find that helical fields, generated by the large-scale dynamo
below, emerge above the solar surface. We expect such
fields to drive flares and coronal mass ejections via Lorentz force. 
\section{The Model}
\label{model}
Similar to Warnecke \& Brandenburg (2010), a two-layer system is used.
We model the convection zone starting at $r=0.7\,R_\odot$ to
the solar corona till $r=2\,R_\odot$, where $R_\odot$ is the solar radius,
used from here on as our unit length.
Our simulation domain is a spherical shell extending 
in the $\theta$ (colatitude) direction from $\pi/3$
to $2\pi/3$ and in the $\phi$ direction from $0$ to
$0.3$. This means, the surface in the photosphere, which will be
described by this simulation, would 600 $\times$ 200 Mm$^2$ large.
A helical random force drives the velocity in the lower layer.
For our model the momentum equation can be written as followed:
\begin{equation}
{\DD\UU\over\DD t}=\theta_w(r)\ff +\nabla h + \gggg
+\JJ\times\BB/\rho+\FF_{\rm visc},
\label{DUDtext}
\end {equation}
where  $\theta_w(r)=\half\left(1-\erf{r\over w}\right)$, a profile
function the connect the two layers, where $w$ is the width of the
transition.
 and $\FF_{\rm visc}=\rho^{-1}\nab\cdot(2\rho\nu\SSSS)$ is the viscous force,
${\mathsf S}_{ij}=\half(U_{i;j}+U_{j;i})-\onethird\delta_{ij}\nab\cdot\UU$
is the traceless rate-of-strain tensor, semi-colons denote covariant
differentiation, $\gggg=-\frac{GM}{r^2}\rrr$ the gravitational acceleration,
$h=\cs^2\ln\rho$ is the specific pseudo-enthalpy,
$\cs=\const$ is the isothermal sound speed,
and $\ff$ is a forcing function that drives turbulence in the
interior.
The pseudo-enthalpy term emerges from the fact that for an isothermal
equation of state the pressure is given by $p=\cs^2\rho$, so the pressure
gradient force is given by $\rho^{-1}\nab p=\cs^2\nab\ln\rho=\nab h$.
The continuity equation be written in terms of $h$
\begin{equation}
{\DD h\over\DD t}=-c_s^2\nab\cdot\UU.
\label{dhdt}
\end{equation}
\EEqs{DUDtext}{dhdt} are solved together with the induction equation.
In order to preserve $\nab\cdot\BB=0$, we write $\BB=\nab\times\AAA$ in terms
of the vector potential $\AAA$ and solve the induction equation in the form
\begin{equation}
{\partial\AAA\over\partial t}=\UU\times\BB+\eta\nabla^2\AAA,
\end{equation}
For the density we use an initial distribution, where $\rho \approx
1/r^2$.

The simulation domain is periodic in the azimuthal direction. 
For the velocity we use the
stress-free conditions all other boundaries.
For the magnetic field we adopt vertical field conditions for the
$r=2$ boundary and perfect conductor conditions for the $r=0.7$ and
both $\theta$ boundaries.
Time is measured in non-dimensional units $\tau = t \urms\kf$, which is
the time normalized to the eddy turnover time of the turbulence.
We use the 
{\sc Pencil Code}\footnote{\texttt{http://pencil-code.googlecode.com}},
which uses a sixth order centered finite-difference in space and 
a third-order Runge-Kutta scheme in time. 
See Mitra,Tavakol, Brandenburg \& Moss (2009) for extension of 
the {\sc Pencil Code} to spherical coordinates. 
\section{Results}
\begin{figure}\begin{center}
\includegraphics[width=3.3cm]{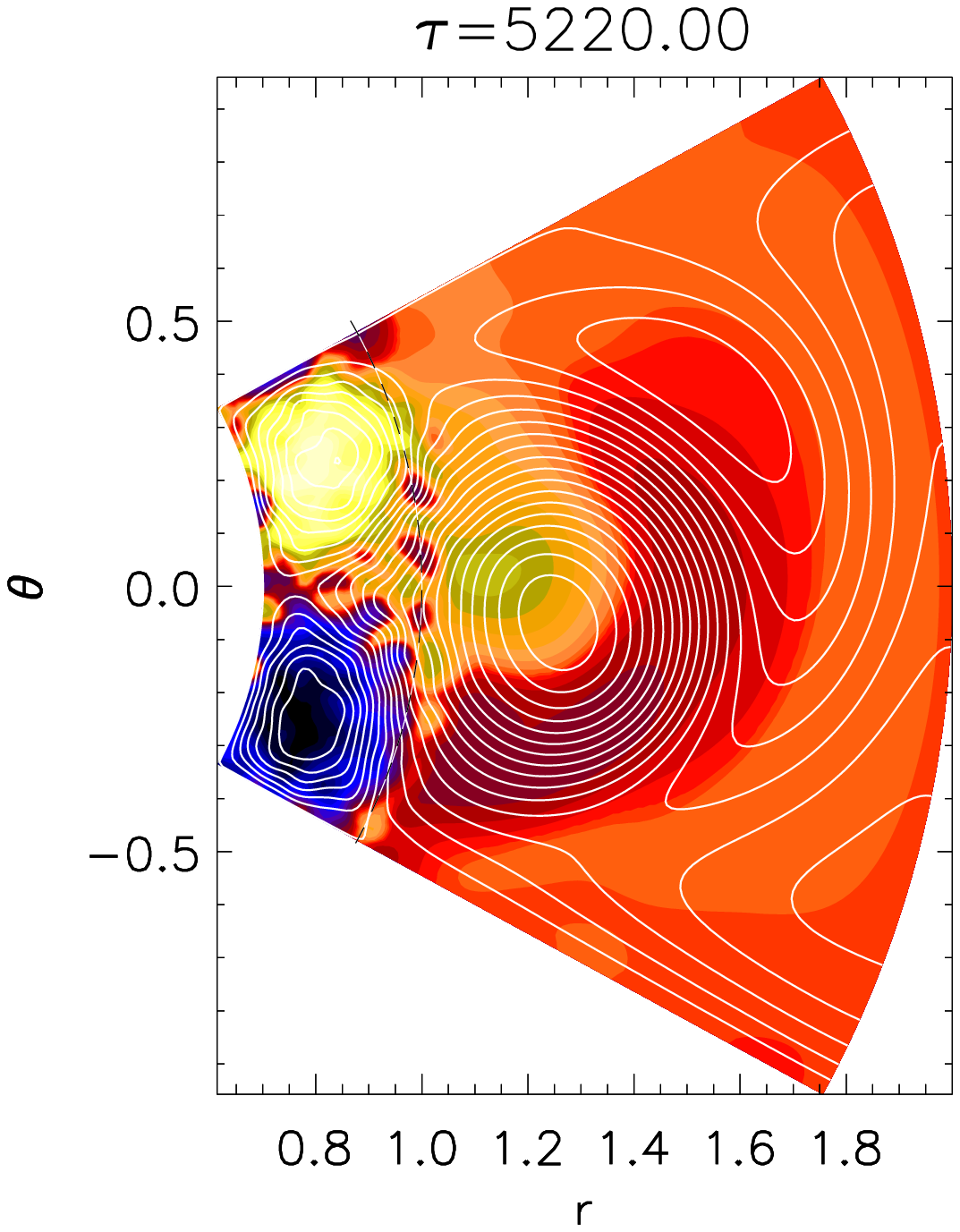}
\includegraphics[width=3.3cm]{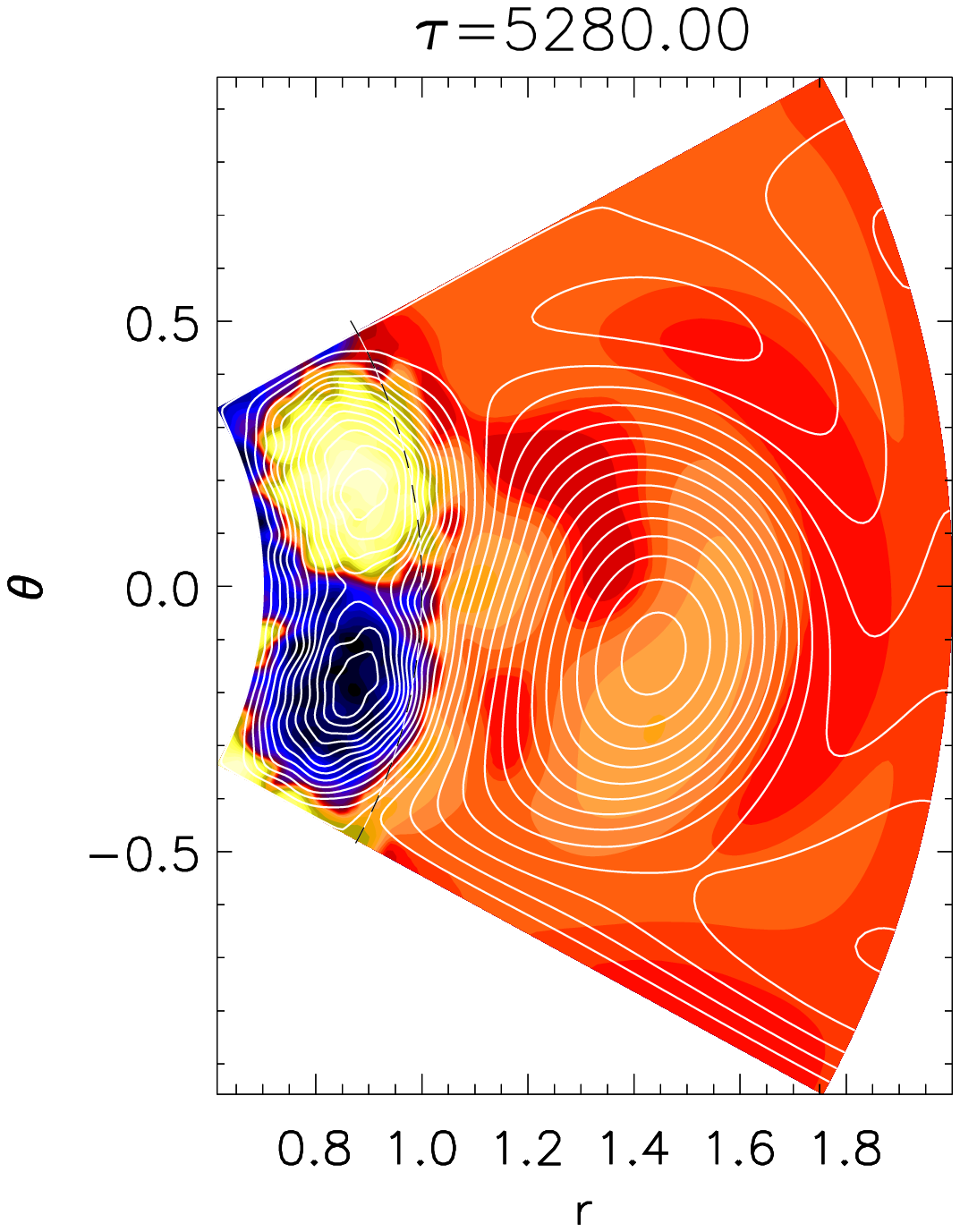}
\includegraphics[width=3.3cm]{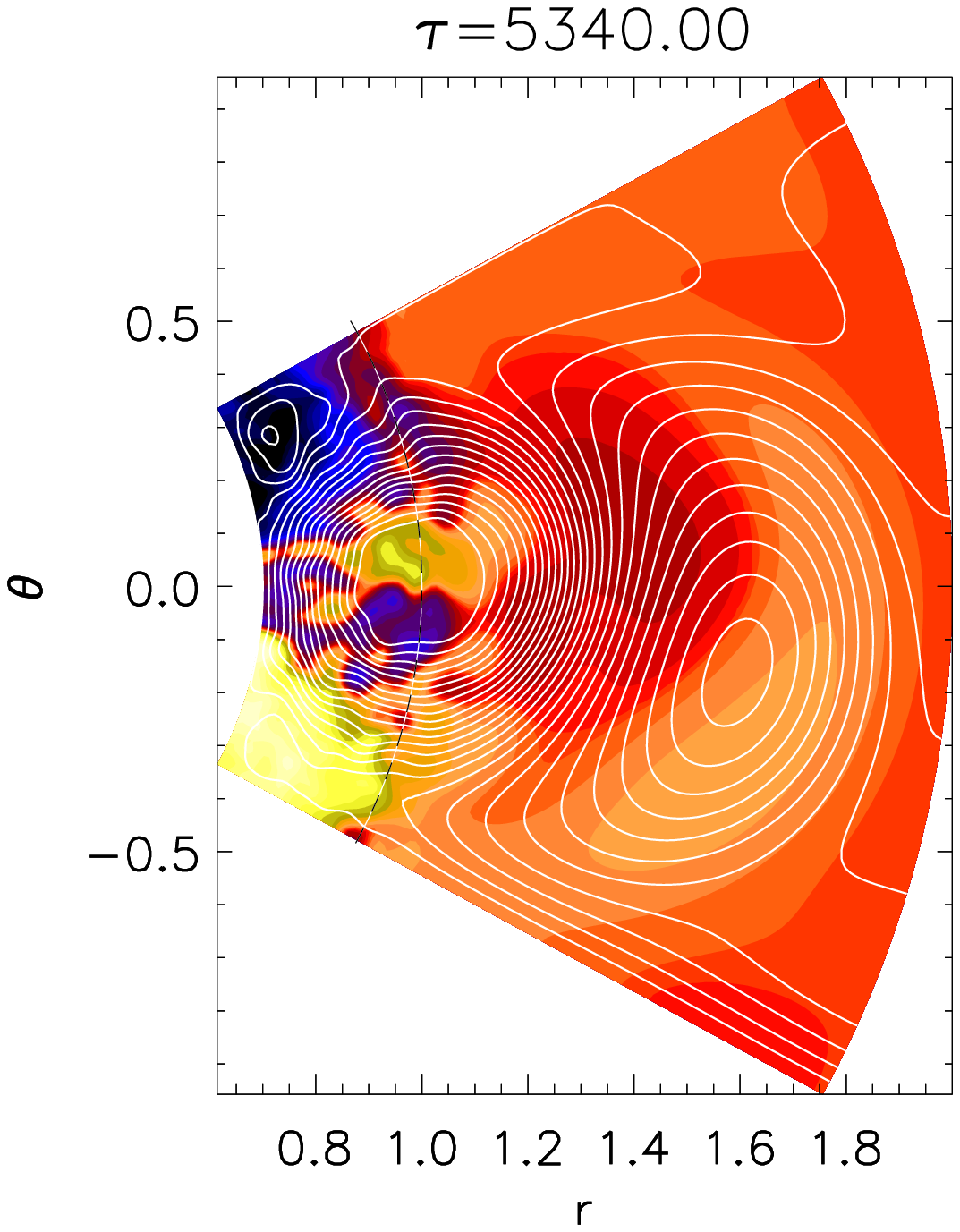}
\includegraphics[width=3.3cm]{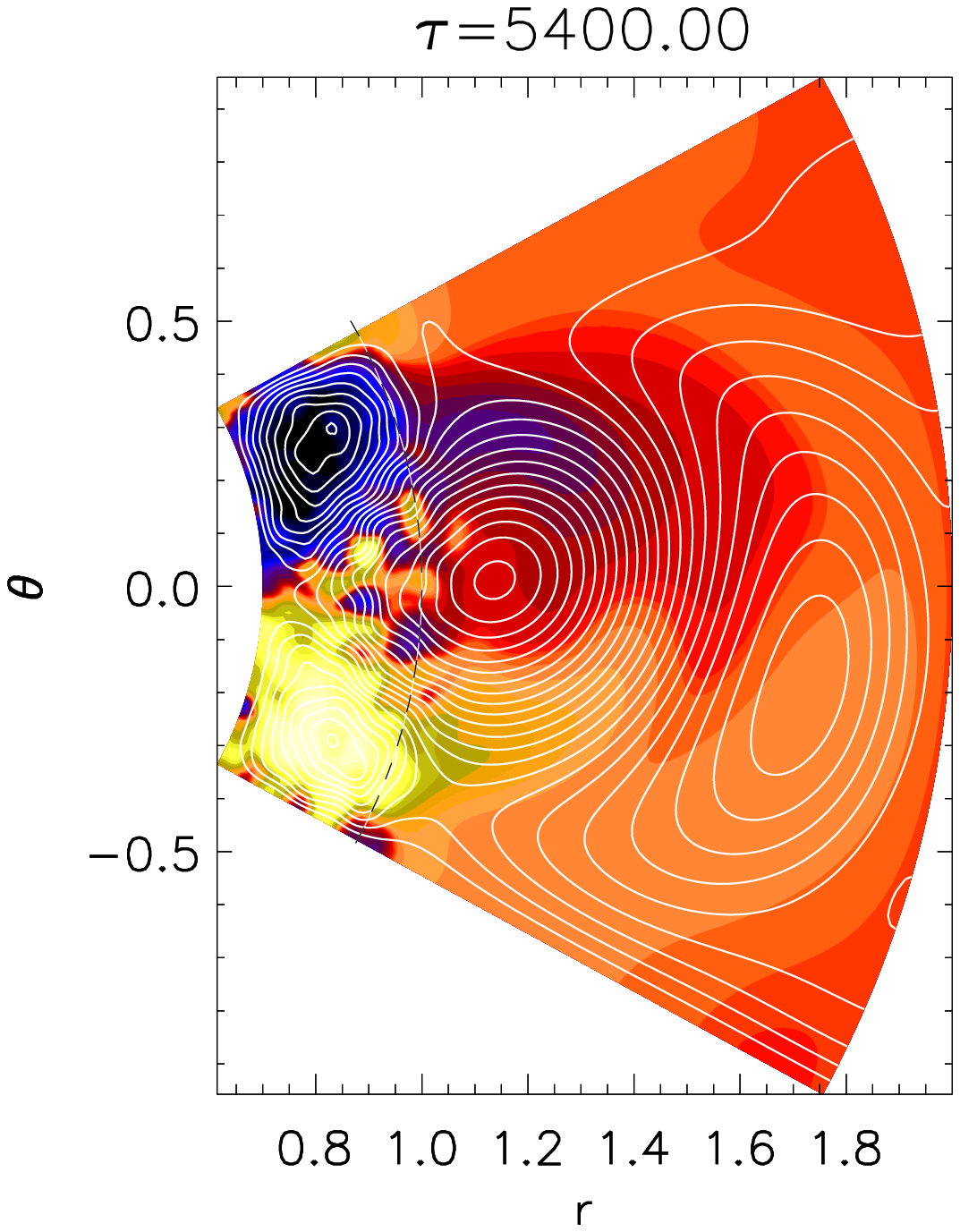}
\end{center}\caption[]{
Time series of formation of a plasmoid ejection in spherical
coordinates. 
Contours of $r\sin{\theta}\bra{A_{\phi}}_{\phi}$ are shown together
with a color-scale representation of $\bra{B_{\phi}}_{\phi}$; dark
blue stands for negative and red for positive values.
The contours of $r\sin{\theta}\bra{A_{\phi}}_{{\phi}}$ correspond to
field lines of $\bra{\BB}_{\phi}$ in the $r\theta$ plane. 
The dotted horizontal lines show the location of the surface at $r=1\,R_\odot$.
}
\label{A_sph}
\end{figure}

\begin{figure}[t!]
\begin{center}
\includegraphics[width=\columnwidth]{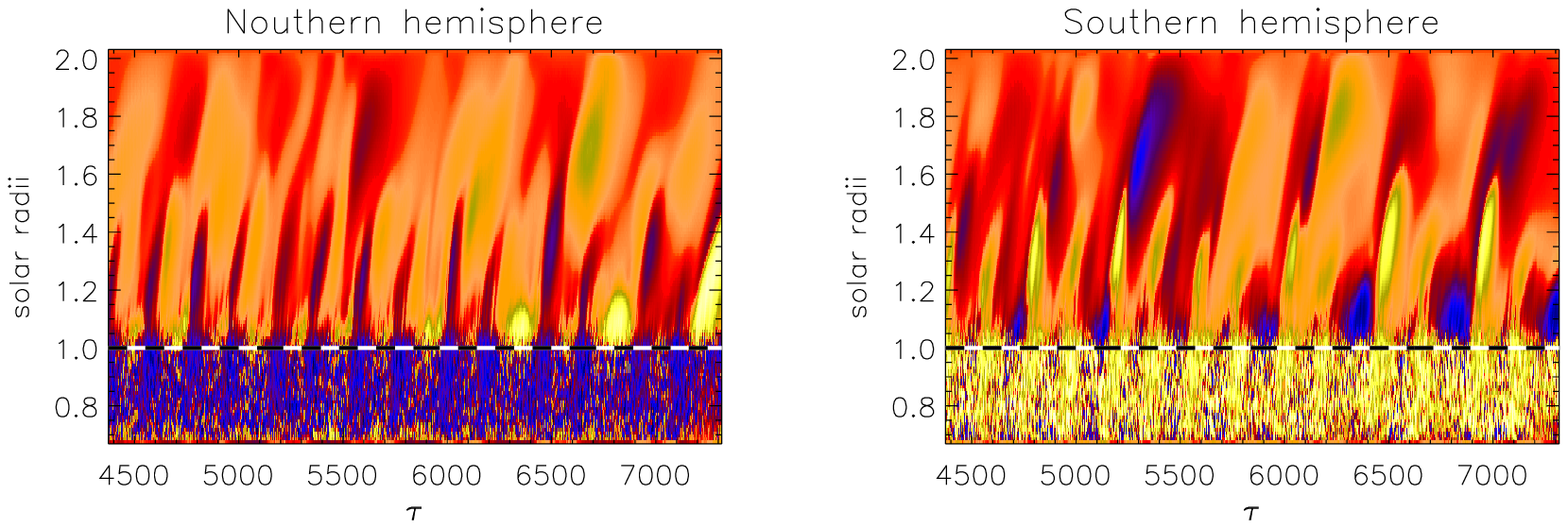}
\end{center}\caption[]{
Dependence of $\bra{{\JJ\cdot\BB}_{\phi}\,\theta} / \overline{\bra{\BB^2}_{\phi
\theta}}$ versus time $\tau$ and radius $r$ in terms of the solar
radius $R_\odot$. 
The left penal show a thin band in $\theta$ in the northern hemisphere,
the right one a thin band in $\theta$ in the
southern hemisphere, both are averaged over 20$^{\circ}$--28$^{\circ}$ latitude.
Dark blue stands for negative and red for positive values.
The dotted horizontal lines show the location of the surface at $r=1\,R_\odot$.
}
\label{jb2}
\end{figure}

\begin{figure}[t!]
\begin{center}
\includegraphics[width=\columnwidth]{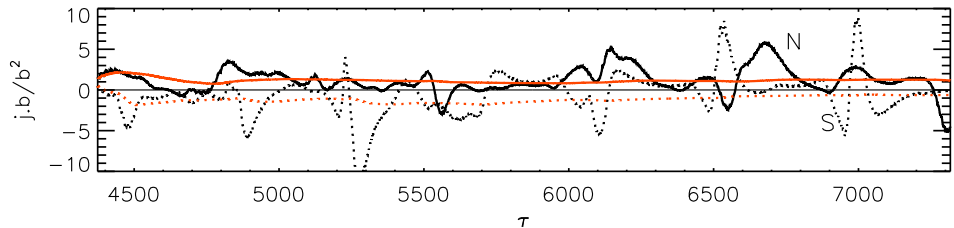}
\end{center}\caption[]{
Dependence of $\bra{\JJ\cdot\BB}_{\phi} /
\overline{\bra{\BB^2}_{\phi}}$ versus time $\tau$ for a radius of
$r=1.5\,R_\odot$ at $\pm28^{\circ}$ latitude in arbitrary units.
The solid lines stand for the northern hemisphere and the dotted for
the southern hemisphere.
The red colored lines represent accumulated means for each hemisphere.
}
\label{jbP}
\end{figure}

The forcing gives rise to an $\alpha^2$ dynamo in the turbulence
zone.
After a short phase of exponential growth, the magnetic field
shows opposite polarities in the two hemispheres with oscillations and
equatorward migration (Mitra et al.\ 2010).
The maximum magnetic field at each hemisphere is 
about 63\% of the equipartition value. 
This is a typical behavior of an efficient large-scale dynamo.
The magnetic fields emerge through the surface and create field line concentrations,
which reconnect, separate and rise to the outer boundary of the domain.
This dynamical evolution is clearly seen in a sequence of field line
images in \Fig{A_sph}, where the field lines of $\bra{\BB}_{\phi}$ are
shown as contours of $r\sin{\theta}\bra{A_{\phi}}_{\phi}$ and the
color representation stands for $\bra{B_{\phi}}_{\phi}$. 
\begin{figure}[t!]\begin{center}
\includegraphics[width=3.3cm]{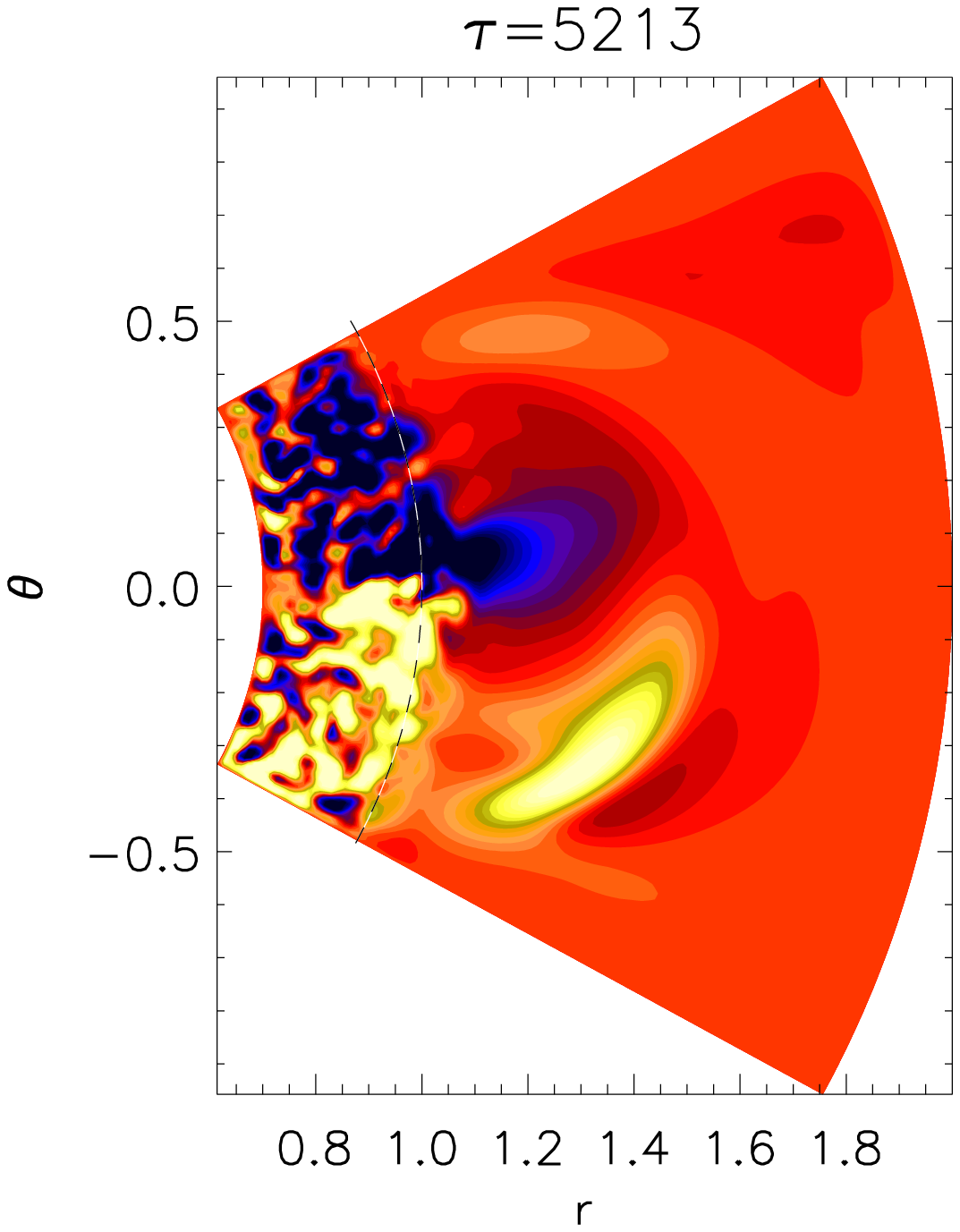}
\includegraphics[width=3.3cm]{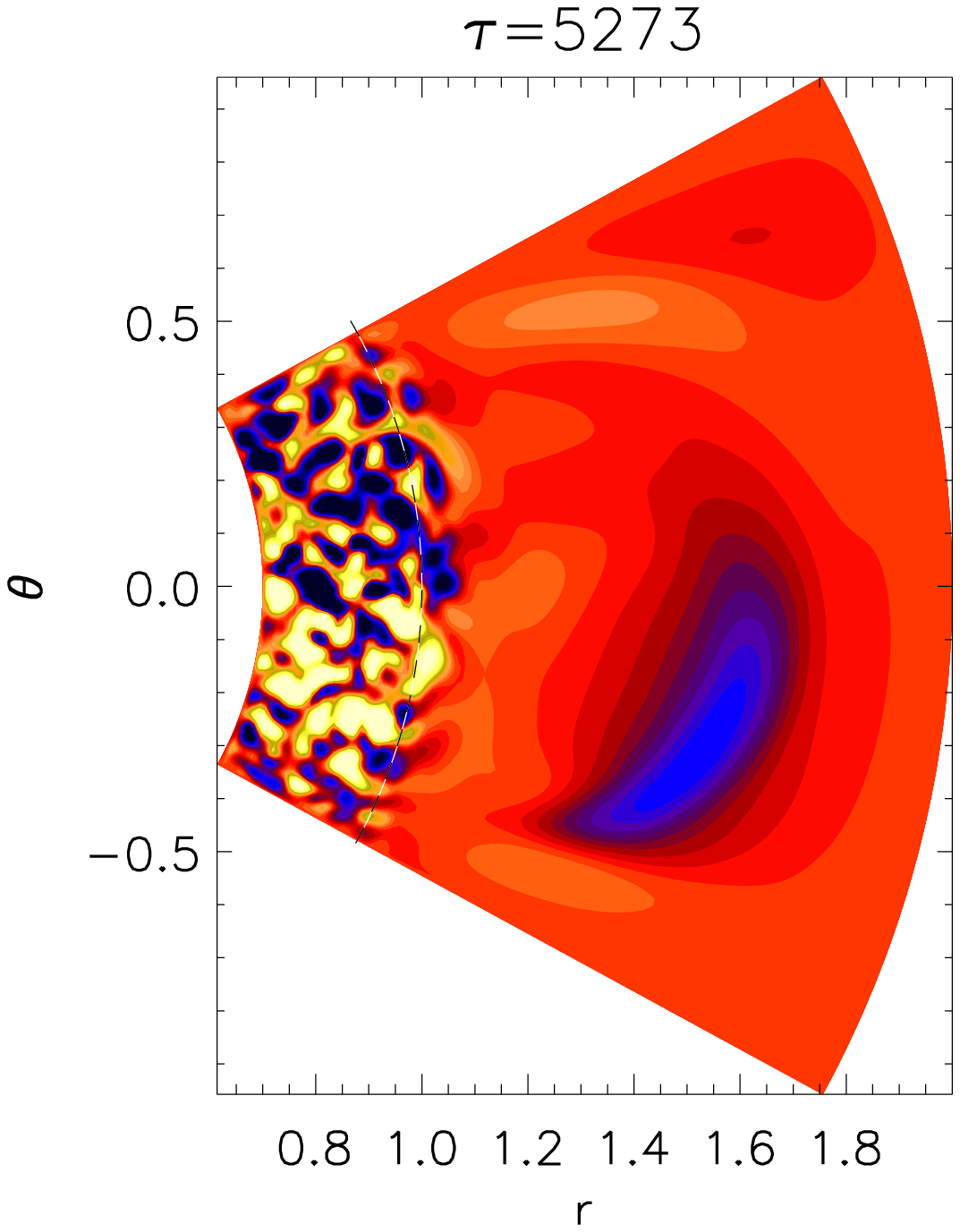}
\includegraphics[width=3.3cm]{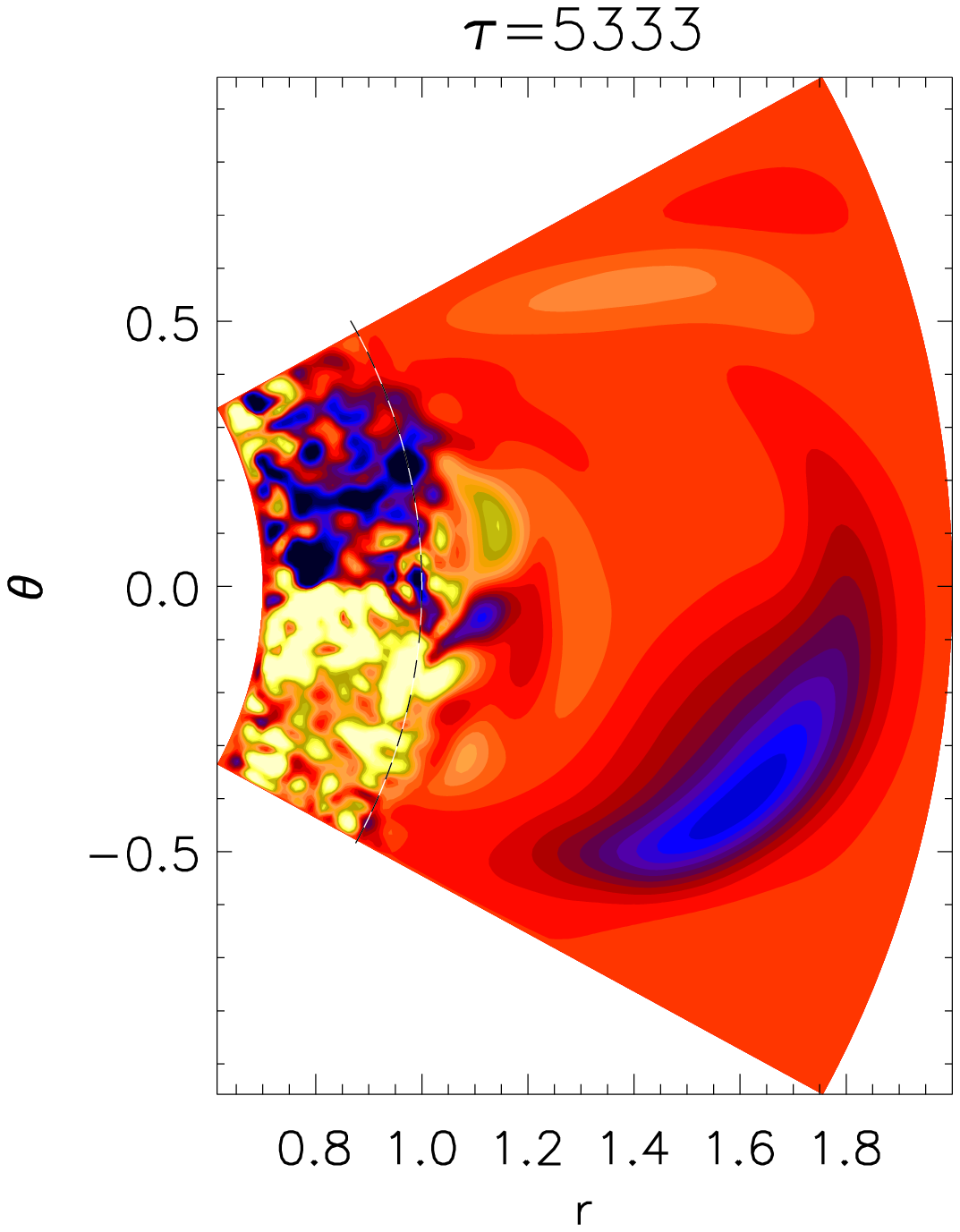}
\includegraphics[width=3.3cm]{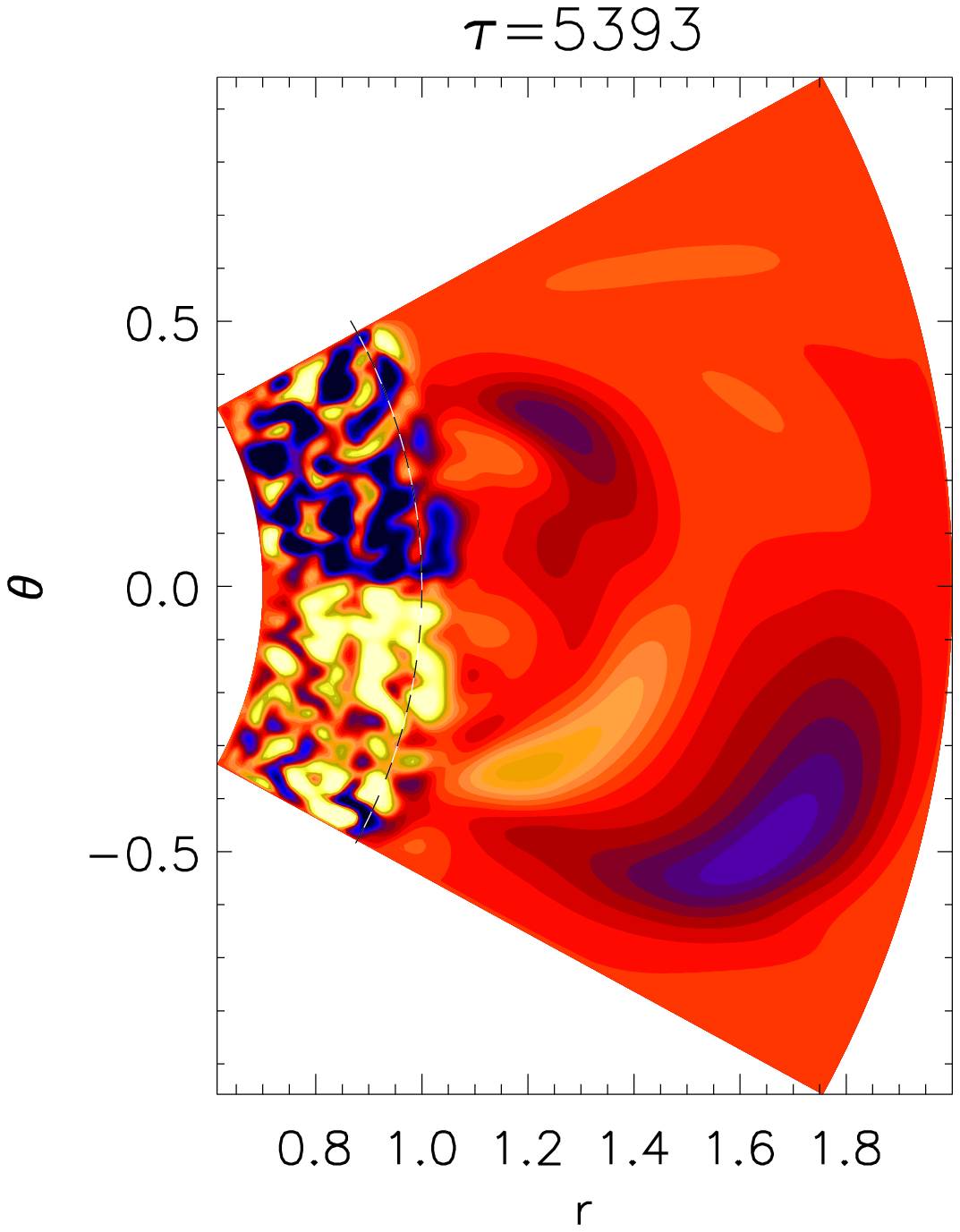}
\end{center}\caption[]{
Times series of what looks a bit like coronal ejections in spherical coordinates.
The normalized current helicity
$\bra{\JJ\cdot\BB}_{\phi}/\overline{\bra{\BB^2}_{\phi}}$ is shown in a
color-scale representation for different times; dark blue stands for
negative and red for positive values.
The dotted horizontal lines show the location of the surface at $r=1\,R_\odot$.
}
\label{jb_sph}
\end{figure}
Investigating the current helicity, we find a surprising result.
In the turbulence zone the  $\bra{\JJ\cdot\BB}_{\phi}/\bra{\BB^2}_{\phi}$
is negative in the northern hemisphere and positive in south, i.e.,
has the same sign as the helicity of the external force.
Above the surface, for each hemisphere, 
current helicity with sign opposite to the turbulent
layer is ejected in large patches (\Fig{jb_sph}).
These structures correlate with reconnection events of strong magnetic
fields.
In Rust (1994) such phenomena have been described as {\it magnetic clouds}.
In order to demonstrate that plasmoid ejection is a recurrent phenomenon,
we plot the evolution of the ratio $\bra{\JJ\cdot\BB}_{\phi}/\bra{\BB^2}_{\phi}$
versus $\tau$ and $r$ in \Fig{jb2}.
We further find that the typical speed of plasmoid ejection
is about 0.13 times the rms velocity of the turbulence in the
interior region, which corresponds to 0.08 of the Alfv\'en speed.
The time interval between successive ejections is about 100 $\tau$.
Note that the ejection of magnetic field and associated reconnection
events are fairly regular, but the formation structures
such as {\tt magnetic clouds} are less regular. 
Note further that the 
sign of azimuthally averaged current helicity in the outer layer is
always opposite to that of the turbulence zone.
This is demonstrated in \Fig{jbP} where we plot time-series of
azimuthally averaged current helicity at $r=1.5\,R_\odot$.
For the northern hemisphere the current helicity (solid black line) and the
accumulated mean (solid red line) show positive values and for the
southern hemisphere (dotted lines) negative values.
This suggests that, even though the plasmoids observed in our simulations
are shedding small-scale current helicity of opposite sign
to that of the large-scale current helicity inside the turbulence zone,
outside the turbulence zone the sign of large-scale current helicity
has reversed and is now the same as that of the small-scale helicity.
This may be explained by the action of turbulent magnetic diffusion
on the magnetic field;
see e.g., Brandenburg, Candelaresi, \& Chatterjee (2009).

In summary, it turns out that  twisted magnetic fields generated
by the helical dynamo beneath a spherical surface are able to produce flux emergence
in ways that are reminiscent of that found in the Sun. 
We find phenomena that can be interpreted as recurrent plasmoid
ejections, which then lead to magnetic clouds further out.
A promising extension of this work would be to include a Parker--like wind 
that turns into a supersonic flow at sufficiently large radii.

\end{document}